# Evaluating energy inefficiency in energy-poor households in India: A frontier analysis approach

Vallary Gupta (1), Ahana Sarkar (2), Chirag Deb (3), Arnab Jana (1)

((1) Department of Civil Engineering, Indian Institute of Technology Bombay, Powai, Mumbai, Maharashtra, India, (2) Indian Institute of Technology Bombay, Powai, Mumbai, Maharashtra, India, (3) School of Architecture, Design and Planning, The University of Sydney, Sydney, Australia)

Energy-poor households often compromise their thermal comfort and refrain from operating mechanical cooling devices to avoid high electricity bills. This is compounded by certain behavioral practices like retention of older, less efficient appliances, resulting in missed energy savings. Thus, the need to enhance efficiency becomes critical in these households. However, due to a lack of comprehensive data in India, little is understood about their electricity consumption patterns and usage efficiency. Estimating inefficiency and assessing its determinants is crucial for improving their quality of life.

This study measures the inefficiency in electricity consumption due to household practices and appliances in social housing in Mumbai, India. It considers technological determinants in addition to socio-economic variables. The study employs primary data collected from rehabilitation housing and slums in Mumbai. Stochastic frontier analysis, a parametric approach, is applied to estimate indicators of electricity consumption and inefficiency. While household size and workforce participation significantly affect consumption behavior in rehabilitation housing, it is limited to the workforce in slums. The ownership of appliances, except for washing machines in slums, also exhibits considerable impacts. The mean efficiency scores of 83% and 91% for rehabilitation housing and slums, respectively, empirically quantify the potential savings achievable. Factors that positively influence inefficiency include the duration of operating refrigerators, washing machines, iron, and AC. These results hold implications for enhancing the uptake of efficient appliances in addition to accelerating energy efficiency retrofits in the region. Policies should focus on awareness and the development of appliance markets through incentives.

**Nomenclature**

***Abbreviations***

**AC** - Air conditioner

**CFL** - Compact fluorescent lamp

**DEA** - Data envelopment analysis

**HH** - Household

**IEA** - International Energy Agency

**IDA** - Index decomposition analysis

**LED** - Light emitting diode

**LIG** - Low- income group

**LR** - Likelihood ratio

**MLE** - Maximum likelihood estimation

**NHN** - Half Normal

**OECD -** Organization for Economic Co-operation and Development

**OLS -** Ordinary least squares

**SFA** - Stochastic frontier analysis

**S.D.** - Standard deviation

**SDG** - Sustainable development goals

**SEEF** - Stochastic energy efficiency frontier

**SRH** - Slum rehabilitation housing

**TV** - Television

**TN** - Truncated normal

**UN-HABITAT** - United Nations Human Settlements Programme

**WFPR** - Workforce participation rate

***Symbols***

$Q$- Electricity demand

$P$- Price

$I$- Income

$Y_i$ - Electricity consumption for household $i$

$X_i$ - Vector of explanatory variables for household $i$

$\beta$ - Coefficient

$v_i$ - Random errors

$\sigma_v^2$ - Variances of random errors

$u_i$ - Non-random error term, or inefficient energy

$\sigma_u^2$ - Variances of inefficiency

$f(Z_i)$ - Variables of household electricity consumption explaining the inefficient energy

$\delta$ - Vector of parameters

$\varepsilon_i$ - Random errors in inefficiency function

$TF_i$ - Technical efficiency

$E_i^F$- Minimum consumption

$Ei$ - Actual consumption

$L(H_o)$ – Log-likelihood value of OLS model

$L(H_1)$ - Log-likelihood value of SFA model

***Subscripts***

$i$ – Household

$u$ – Random error component

$v$ – Inefficiency component

## 1. Introduction

Energy access is critical for reliable healthcare, lighting, heating, cooking, mobility, and telecommunications services [1]. Greater and secure access to energy is often associated with improved thermal comfort, which correlates to better productivity, health, and well-being [2]. Ensuring access to affordable, efficient, reliable, and sustainable energy has also been one of the 17 United Nations Sustainable Development Goals (SDGs). It is also critical to eliminate poverty (SDG 1), health hazards (SDG 5), and climate change (SDG 13).

The building sector globally accounts for more than one-third of energy demand and more than 35% of energy-related carbon emissions [3]. Building operations alone consume more than 55% of global electricity [4]. According to the International Energy Agency (IEA), finding energy-efficient solutions is essential to reducing the rising energy demand in the building sector, which is projected to increase by 30% by 2060. United Nations Framework Convention on Climate Change (UNFCC) also underscores the importance of addressing energy efficiency in buildings for emission reductions to achieve the Paris Agreement goals. Identifying the potential steps for achieving energy efficiency is vital for reducing overall energy consumption; this calls for reliable estimates of the existing inefficiencies in energy use.

In India, 30% of the total energy consumption is used by the building sector, involving cooking, lighting, water heating, ventilation, refrigeration, and air conditioning [5]. In the face of rising urbanization, increasing floor space, and growth in air conditioners (ACs) and appliances, this share is expected to grow in the coming decades [3]. Energy demand in India is expected to double by 2040, with the electricity demand potentially tripling due to rising cooling needs and appliance ownership. Therefore, building energy efficiency improvements offers enormous potential for social, economic, and environmental benefits. Substantial advances in energy efficiency are critical to avoid massive national investments.

Energy inefficiency in the residential sector arises primarily due to the presence of energy-inefficient technologies or inefficient use owing to behavioural patterns. Varying patterns of socioeconomic background, demographic characteristics, and appliance ownership affect behavioral patterns and, consequently, efficient electricity consumption [6,7]. While appliances are termed as direct determinants of building energy consumption, others are referred to as indirect determinants that affect heat gains/ losses, lighting, etc., and thus energy consumption by influencing the ownership and use of appliances. Literature reports inefficiencies ranging between 20 to 27% due to household characteristics and appliance usage patterns [8]. A recent study shows that low-income households exhibit lower efficiency scores compared to wealthier households, owing to the latter's ability to afford more efficient appliances and a higher willingness to pay for energy efficiency [9]. Energy demand triggered by affordability and certain adaptive practices can affect consumption in these households [10]. Gender differences may also affect efficient consumption. Women are often less involved in decision-making and exhibit lesser awareness of energy-efficient upgrades and schemes [11]. Households with a male household head have been found to use electricity more efficiently [9]. A recent study in India suggests that the type of appliance ownership (purchased new, second-hand, or bought from a relative/ friend), higher initial capital costs, efficiency levels or star labelling, electricity bills, and maintenance also affect the efficiency [12]. The age of the appliance has an effect in the range of 28% to 36% [13].

Numerous studies have estimated energy efficiency in regional and household or firm levels [9,14]. Inefficiency estimates have been calculated for countries like the US, Canada, Australia, the EU, Japan and India, including OECD nations [14–17]. However, little attention is paid to India's residential energy efficiency, in particular, the energy-poor families using household-level estimates. Research suggests that people in low-income strata, driven by personal aspirations and poor thermal comfort [18], tend to keep less efficient household appliances and equipment due to financial constraints [19]. Unlike higher-income populations who report higher electricity use efficiency [16], these households have the least prevalence of Energy Star appliances [20]. They are more likely to have older appliances that are less frequently retired from use, purchased from a second-hand market, or obtained at no cost from friends and family [21]. High up-front costs often refrain them from making energy efficiency upgrades. Subsequently, they usually compromise their thermal comfort to avoid financial stress due to higher electricity bills and a poorly built environment [18]. Residents often restrict their use of mechanical cooling devices [22], and resort to natural ventilation or personal adjustments to feel comfortable indoors [23]. Recent evidence suggests that appliance diffusion can remain low despite rising incomes if appliances are too expensive [24]. Additionally, these households spend on consumption upgrades such as new purchases only after fulfilling their survival needs [25].

Thus, opportunities to save electricity in the residential sector are significantly linked to the preferences and decisions in choosing efficient equipment, energy-saving practices, and the level of awareness and the age of installed appliances [26]. Families that are poor or are a part of a second-hand culture often report higher electricity consumption and subsequent reduction in thermal comfort due to compromised hours of usage [23]. While higher appliance ownership is linked to improved living standards, a higher degree of indoor comfort and convenience [27], energy-poor households in India often compromise their comfort, are part of specific behavioral and cultural practices, and use second-hand or obsolete appliances to reduce financial burdens. Despite the long-term economic advantages surpassing the initial expenditure [28], these households often miss the potential for energy savings and witness adverse effects on health. Additionally, as exposure to internal overheating due to unmet cooling needs, poor envelope design, and indoor environmental conditions can lead to adverse health risks, the opportunity and the need to enhance efficiency is critical in these households. This will be further aggravated by increased events of extreme heat and cold due to climate change in the coming decades [29].

An estimated 30-50% of the total population in developing countries like India is defined as urban poor [30]. They often lead a restricted living owing to financial constraints. A lack of comprehensive data renders the strategies to enhance energy efficiency, particularly for such energy-poor families residing in social housing like slums and SRH, uninformed. Gaining insight into their electricity consumption patterns and determinants of inefficiency is critical for improving their overall quality of life. The present literature offers an extensive outline of appliance ownership and its effect on energy consumption. Considerable efforts have been made to understand the factors affecting appliance ownership and household electricity consumption. As these households shift from horizontal slums to vertical SRH, with the security of tenure, they are more likely to indulge in appliance purchases, in particular, older ones. However, the inefficiencies that result from these household patterns often remain under-studied. Also, several studies have examined the effects of labelling [31], knowledge and awareness [32], technology [33], pricing policy, and temporal preferences [34] on the uptake of energy-efficient appliances. However, limited attempts have been made to evaluate the existing inefficiencies in older appliances, particularly in low-income groups.

Further, while inefficiencies have been evaluated in economically stable households to devise policy measures for guiding behavioral changes that curb inefficient use, the trends have been less explored for energy-poor households, particularly in India. While literature evidence that the shift from informal slums to formal rehabilitation units yields variations in social aspiration, appliance purchase decisions, and usage behavior, its subsequent impact on inefficiencies is limited. A study by Blasch et al. suggests that the energy consumption for providing a given level of energy service can be reduced either by improving the efficiency levels of using inputs (i.e., appliances) or by adopting new energy-saving technology (i.e., purchasing new appliances or opting for energy-saving renovations), or both [35]. The reduction in residential energy consumption is, thus, a result of the interaction between changes in technology and household behaviour [36].

A large body of existing research employs regression analysis to empirically predict determinants of efficiency in energy consumption and uptake of efficient appliances [20,37]. Index Decomposition Analysis (IDA) has also been employed to estimate savings due to efficiency improvements [15,38]. However, it is pointed out that the efficiency savings from IDA estimates differ significantly from the government’s estimates [15]. To overcome the shortcomings, non-parametric technique—Data Envelopment Analysis (DEA) [39,40] and parametric-based techniques, Stochastic Frontier Analysis (SFA) [14,16,41,42] have been applied to estimate energy efficiency. These benchmarking models compare a building or an observation with the best-performing observation (or best-practice technology) in its own class, allowing for assessing the total energy efficiency potential. The DEA uses a deterministic function where no specific functional form is necessary. However, in the parametric approach, the SFA functional form is imposed. Unlike DEA, SFA model can separate actual output deviating from a frontier into inefficiencies and noise [17]. Additionally, it has a stronger theoretical foundation and a higher discriminating power that aids in overcoming measurement problems [43]. SFA model is thus found more suitable for estimating the efficiency level in electricity use.

The current study examines both the effect of household behaviour and technology (current appliance stock) in energy-poor households for a selected case study area in India. Using the parametric approach, stochastic frontier energy demand model, this study intends to assess space electricity consumption efficiency in households in slums and SRH in Mumbai, India, by conducting a household survey. The objectives of this study are to: (1) disaggregate indoor residential electricity consumption into the minimum demand (frontier and over-consumption) in slums and SRH, (2) determine the key factors, income and non-income drivers, influencing the electricity consumption and inefficiency of electricity usage, and (3) to empirically quantify potential electricity savings or usage efficiency in electricity consumption.

The significant contribution of this study lies in estimating the inefficiencies in electricity consumption caused by household characteristics and user behaviour in terms of appliance usage patterns for slums and SRH in India. The research hypothesizes that household characteristics, appliance ownership, and usage patterns result in inefficiencies in energy-poor households. It further tests the hypothesis that SRH experiences higher inefficiencies than slums. The significance of this study lies in providing a foundational framework for designing programs to augment access to vital energy services such as cooling and health preservation. They are critical in creating incentives and advocating policy changes to adopt energy-efficient practices in social housing. Improving electricity efficiency and associated awareness in these households is crucial in developing a more equitable, healthy, and energy-sustainable community.

## 2. Theory

Regression analysis is the most widely used method to empirically predict energy consumption using measured data on energy usage and variables such as climate, building design and characteristics, appliances and equipment, and occupant behaviour [44]. This method identifies the average trend over the whole sample population and compares each observation with that overall trend. However, the analysis may result in a poor interpretation of poorly performing observations by skewing the regression line toward less efficient samples [40]. Economic approaches such as the classic nonparametric DEA [39] and the parametric stochastic frontier analysis [41] overcomes these shortcomings by using efficiency analysis.

The first-ever estimates of energy demand were performed by Houthakker [45] using the classical demand theory followed by Griffin [46] who estimated energy demand using income and price. This was followed by Halvorsen, who again emphasized estimates of price and income elasticities [47]. Other studies examined factors like demographic features [48], appliance stock and prices [49]. It was later that Hartman discussed the importance of efficiency [50]. However, these studies did not empirically quantify the efficiency in the consumption of energy. Studies started quantifying efficiency, predominantly in the context of a production function, only after Aigner et al. outlined the stochastic frontier analysis method [41]. However, the approach is focused on behaviour of firms, and not the household sector. Attempts to measure energy consumption efficiency at the level of individual households were made recently by Filippini and Hunt [51,52], owing to the limited availability of methods and techniques in addition to data. The authors provided an alternative view of energy efficiency based on the aggregate energy demand function or a ‘frontier energy demand function'. The function extends the traditional econometric/empirical demand functions to consider a household’s desire to minimize its consumption and subsequent cost. The demand models define the minimum level of energy that a household can reasonably consume using given resources. It was later that Broadstock et al., [53] measured energy efficiency in Chinese households using the concept of production efficiency introduced by Filippini and Hunt.

SFA has been widely used to evaluate energy efficiency in buildings, vehicles, industries, and households since then [54,55]. The model decompose the error term into a random error that captures the random effects outside the control and the efficiency component [17]. The frontier assumes that while some deviations can be attributed to random events, others are due to household-specific inefficiency [56]. Filippini and Hunt first examined the energy efficiency in energy demand in OECD countries using panel data via the energy demand SFA approach. They found that energy efficiency depicts a negative association with residential energy consumption intensity [51]. Another study in the USA investigated energy efficiency in households and found that inefficiencies in usage hold a share of 10-17% of energy consumption [57]. Kavousian et al. modified the non-parametric, DEA method and proposed a new statistical method called stochastic energy efficiency frontier (SEEF) to estimate the energy efficiency scores in residential buildings in the United States [40]. Boogen employed a sub-vector input distance frontier function, a modified SFA model function for households in Switzerland [8]. The author calculated inefficiency in electricity use in the range of around 20 to 25% based on inputs like the aggregate stock of appliances, and household characteristics. The authors suggest that inefficiencies can be avoided through behavioral changes like switching off the lights.

Other studies have used SFA to separately model the frontier function and inefficiencies. Nsangou et al. investigated and compared the efficiency in Cameroon’s rural and urban households [14]. The predicted technical efficiency was higher in rural areas (73%) compared to 62% in urban areas. It was also found that household income and awareness of energy conservation practices are positively related to technical efficiency. The latter was the most critical factor in determining technical efficiency. Twerefou and Abeney estimated energy efficiency for Ghana and found the mean efficiency score to be similar (63%) for urban and rural households [55]. While variables like appliance ownership, income, location, and house size were included in the demand frontier, factors like household characteristics, dwelling type, and employment status were used to determine their effect on inefficiency in consumption patterns. The authors conclude that providing energy-efficient appliances and education in rural areas can enhance electricity consumption efficiency. A recent study by Zheng et al., also estimated inefficiencies using the frontier function and an energy efficiency part [25]. The study evaluated the effects of income and other attributes on households in China for three income clusters and found a negative association between income and efficiency. Low-income households showed highest energy efficiency, decreasing slowly, while high-income groups exhibited the lowest efficiency levels, decreasing faster. Growing survival needs (cooking and heating/ cooling) with rising income results in declining energy efficiency in the low-income groups. Unlike the other two income strata, the group reportedly spends on consumption upgrading such as new purchases only after fulfilling their survival needs. For social housing in Mumbai, Gupta et al., and Sarkar et al. point out that most of the appliances present in these households are acquired via donation, resulting in inefficiencies and increased electricity and maintenance costs [23,58]. However, Marin et al., show that appropriate energy efficiency policies and innovation together may aid in achieving energy efficiency targets. The study stresses the effect of energy-efficient technological upgrades and appliance design innovations on efficiency improvements [56].

Andor et al. used disaggregated household data to estimate input requirement function and inefficiencies for 2000 German households. In addition to estimating the models that consider factors such as appliance ownership and usage on the frontier, the study explicitly models the inefficiency determinants, like socioeconomic characteristics such as income, household size, age, etc. The results report a mean inefficiency of around 20%, indicating a notable potential for energy savings [9]. Wang et al. used Chinese household survey data to examine the efficiency of space heating consumption in China and found the average efficiency to be around 69%. The study incorporated factors like household size, income, building types, heating type, etc., on energy demand function and evaluated the impacts of energy efficiency standards, separately on inefficiency function [59]. A recent study from Slovenia also employed disaggregated data from 6882 Slovenian households from two cross-sectional surveys [42]. It considers all types of residential energy fuels and several household factors like dwelling area, household size, number of appliances, the yearly amount of these appliances, and total monthly electricity consumption to measure the inefficiency and identify underlying causes. The authors found considerable heterogeneity in efficiency estimates (varying between 20% to 95%) and attributed the resulting inefficiency to different behavioural patterns and households' energy appliance stock. Weyman-Jones et al. evaluated the efficiency in electricity consumption using time-series data in Portuguese households and estimated the efficiency to be around 60% [60]. The results highlight the success of efficiency measures introduced by the government.

Most of these studies confirm that household income is one of the prominent factors in fostering household energy consumption [61]. However, its effect on energy efficiency is still unclear, particularly in India's energy-poor families. A study in Gujarat, India, suggests a noteworthy reduction in average electricity consumption due to efficient appliances, particularly due to the use of efficient ACs and ceiling fans [62]. T. Zhou & McNeil compared the labeling programs in different economies, including India. The authors employed appliance market share data to quantitatively evaluate the effectiveness of the mandatory programs for appliances such as refrigerators, dishwashers, washing machines, and clothes dryers [63]. Singh et al. used an input-output lifecycle assessment modelling approach. The study concluded that replacing less efficient appliances in India with the best available technologies can reduce overall energy consumption by up to 326 TWh in 2030. This corresponds to a cumulative electricity savings of 937,349 million dollar in 2030 [64]. A similar attempt using a bottom-up energy model highlights that super-efficient appliances, refrigerators, television, ACs, and ceiling fans, can save three times more electricity than India's current standard and labelling program. They can produce an estimated total savings of 60 TWh, and avoid $CO_2$ equivalent of 48 million tonnes [65].

Most of the existing literature stresses the key role of efficient appliances in reducing the rise in energy consumption and GHG emissions if more consumers are motivated by awareness and options. Using national sample survey data (urban and rural), a study found potential household savings in electricity consumption from refrigerators, TVs, ACs, and ceiling fans in the range of 10–27% in 2030 [66]. Though households are inclined to use electricity-efficient star-rated products as per India's Bureau of Energy Efficiency, their effectiveness can be increased by designing labels specific to different states. Additionally, the rapid urbanization and the fast-growing middle-class segment will likely have substantial implications for energy demand, particularly in the residential sector. There is a critical need to manage this demand through energy efficiency improvement [67].

A summary of relevant studies analysing the inefficiencies and their drivers is presented in Table 1. The literature thus reveals a shortage of knowledge on residential electricity consumption behaviours and associated inefficiencies in India, particularly in social housing. Limited attempts have been made to evaluate existing inefficiencies, predominantly in appliance usage, using the energy demand frontier approach in the country.

*Table 1: Summary of existing studies on inefficiencies in energy consumption and its determinants*

| Source | Country | Data source | Independent variables | Inefficiency factors | SFA function |
|---|---|---|---|---|---|
| [25] | China | Government portal | • Energy price<br>• HH income<br>• HH size<br>• Education level<br>• Population | | Energy demand frontier approach with SFA by Filippini and Hunt, 2011. Developed different models to determine distribution of inefficient energy use |
| [59] | China | Government portal | • HH size<br>• No. of elderly<br>• Annual income<br>• House area<br>• Building type<br>• Heating type | • Building envelope performance | Energy demand frontier approach with SFA by Filippini and Hunt, 2011. |
| [42] | Slovenia | | No. of appliances | • Dwelling characteristics<br>• HH characteristics | Shephard energy input distance function. |
| [16] | China | Survey | • HH size<br>• Income<br>• Floor area<br>• Dwelling type<br>• No. of AC and fans | • Income<br>• Building energy efficiency standards<br>• Energy efficiency label<br>• Status of windows and door<br>• Temperature consciousness | Energy demand frontier approach with SFA by Filippini and Hunt, 2011. |
| [9] | Germany | Survey | • Appliance ownership/ use | • Income<br>• Household size<br>• Number of children<br>• Age of the house<br>• Type of family home | Input requirement frontier function by Filippini and Hunt, 2011. |
| [55] | Ghana | | • Price<br>• Income<br>• HH size<br>• Appliance ownership<br>• Ecological zone<br>• Location | • HH head's education<br>• Age of head of HH<br>• Avg. hours of electricity availability<br>• Dwelling type<br>• Poverty status Employment | Energy demand frontier approach with SFA method by Filippini and Hunt, 2011. |
| [14] | Cameroon | | • ICT<br>• Total lamps<br>• Hours of refrigeration<br>• Housing ownership<br>• No. of rooms<br>• HH Size<br>• Type of wall, floor, and roof<br>• House type<br>• Presence of AC, washing machine, computer, TV, fridge | • Income<br>• HH Head gender<br>• Age of head of HH<br>• HH head's education<br>• Mean age of appliances<br>• Employment status,<br>• Favourite brand of appliances<br>• Energy efficiency information on appliances<br>• Energy efficiency awareness<br>• Presence of children under two years of age | Battese and coelli truncated the normal model |
| [68] | China | Government portal | • Floor area<br>• Dwelling type<br>• HH size,<br>• No. of TV and fridge | • Electricity price<br>• Head education<br>• Income<br>• TV labelling tier<br>• Fridge labelling | Stochastic cost function by Aigner et al. (1977) |

## 3. Materials and Methods

A systematic approach based on primary household data, taking Mumbai slums and SRHs as typical case examples of social housing. The research design aimed at enhancing thermal comfort and overall quality of life in social housing adopted in the study is shown in Figure 1. A similar methodology has been followed in recent studies [18]. This study assesses the effect of occupant behavior and technology changes on electricity consumption efficiency. Literature suggests that several factors, including socio-economic characteristics, type of appliances owned (new/ old), and usage behaviour, affect energy consumption and, consequently, its efficiency. In addition to the literature, anecdotal evidence suggests that low-income households compromise their thermal comfort to reduce electricity costs. The study thus assesses the effect of different factors on inefficiency through empirical evidence.

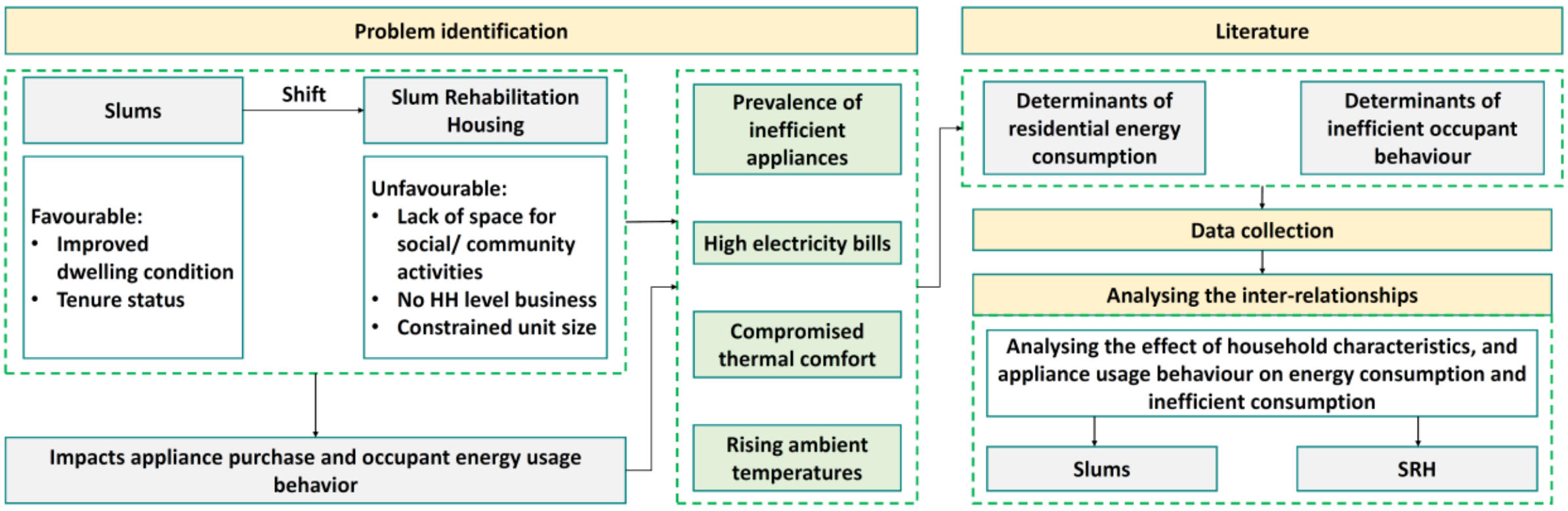

*Figure 1: Methodology adopted in this study*

### 3.1 Theoretical framework and empirical model

The residential demand for energy stems from the demand for energy services such as a cool home, hot water, washed clothes, and can be specified using the basic framework of household production theory. According to this theory, households act as both producing and consuming units. They purchase "goods" which serve as inputs that are used to produce the "commodities". The theory describes the residential demand for energy as a set of inputs that are combined with capital stock to produce energy services that represent outputs. These inputs are energy coming from multiple energy sources used in households (electricity, wooden biomass, natural gas, heating oil, district heat, etc.), that is combined with capital (electrical appliances, heating systems, cooling systems, etc.) to enable households to produce the desired electricity services such as heating or cooling [69].

Following the neoclassical approach [39], it is assumed that for the production of a given level of energy services, households minimize the use of inputs (electricity consumption in this study) and choose the input combination that minimizes production costs. In this way, a household's production function for electricity services is constructed to derive the optimal input demand function. Eq. 1. demonstrates a generalized electricity demand frontier function, where electricity demand (Q) depends on price (P), income (I), and other factors (Z). Q represents the required minimum electricity. and $\epsilon_i$ is the composite error term.

$$Q = f\,(I, Z, P)\; + \epsilon_i \tag{1}$$

The notion of frontier energy demand defines 'frontier' as the minimum level of energy that a household can reasonably consume. The difference between actual and estimated frontier consumption (necessary consumption) could be wasted energy or technology inefficiency [52]. This is illustrated in Figure 2. Point A, lying on the frontier, depicts the most efficient household. However, the actual consumption of a household with similar traits may be at a different point, A*. The distance between point A and A* exhibits excess or over-consumed electricity and indicates inefficiency or an inefficient household A*. The dotted line MER corresponds to the minimum energy consumption.

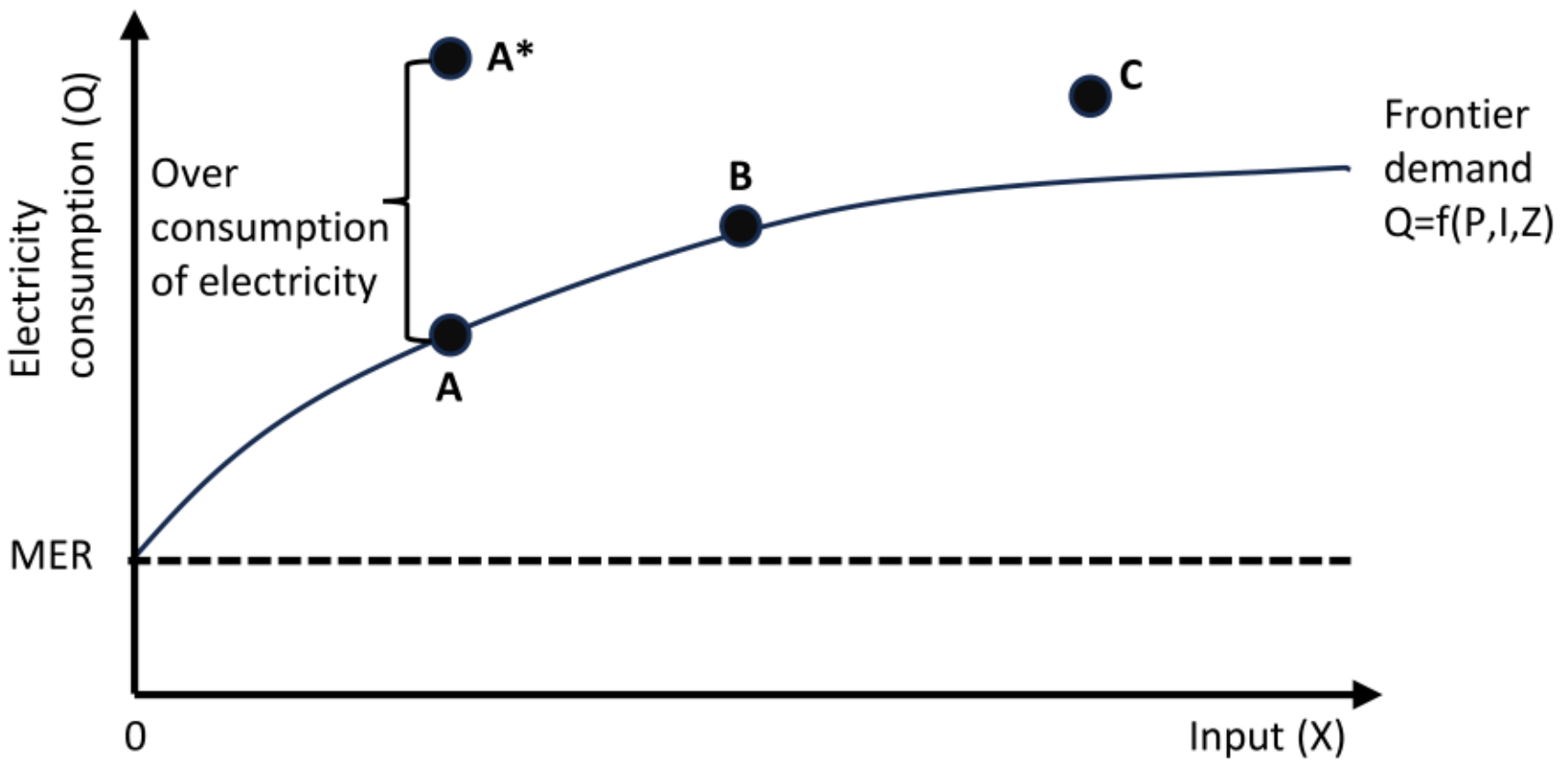

*Figure 2: An illustration of MER and energy demand frontier adopted from [53]*

Filippini & Hunt extended the model in Eq. 1, to include a variable $u_i$ that captures the level of inefficiency, resulting in a stochastic demand frontier presented in Eq. 2 [51]. In the case of an efficient electricity usage process, the error term ($\epsilon_i$) is Gaussian white noise with no wasted electricity. However, since this is unlikely due to the effect of several external factors, $\varepsilon_i$ is segmented into two ($\epsilon_i = v_i + u_i$). Here, $v_i$ is the usual Gaussian white noise. It represents random errors, such as deviation in the data collection process (like the error term in an ordinary least square model), and is assumed to be independently and identically distributed with 0 mean and constant variance of $\sigma_v^2$ [51,70]. $u_i$ represents the non-random error term or the proxy for inefficient electricity. It is the non-random variation related to a household's electricity usage patterns as not all households can use electricity efficiently, and some inefficient use always exists. It is assumed to follow a normal distribution with a constant variance of $\sigma_u^2$. It measures the extra electricity a household uses or the amount that exceeds the frontier demand. As not all households utilise electricity efficiently, and some inefficiencies are always present, $u_i$ follows a one-sided or non-negative distribution, such as the half-normal [41], truncated normal, etc. The inefficiencies can be obtained by following any one distribution.

$$Y_i = f(X_i, \beta) + v_i + u_i$$
$$v_i \sim \mathrm{N} + (0, \sigma_v^2)$$
$$u_i \sim N^+(0, \sigma_u^2) \quad (2)$$

The frontier in Eq. 2 represents the minimum electricity a household consumes to fulfil all the demands in an energy demand function of a household. It shows the optimal electricity usage level when all processes are efficient. electricity efficiency generates the same amount of energy services or outputs with fewer energy inputs. $Y_i$ represents the electricity consumption for household $i$, $X_i$ is the vector of explanatory variables affecting electricity consumption (like income, prize, appliance ownership, etc.) and $\beta$ is an unknown parameter.

The energy demand frontier model, as proposed by Filippini & Hunt, delineates the minimal energy consumption of a family based on its capital stock and several household attributes, including income, age, number of appliances, household size [51]. Any deviation from this minimum energy results in inefficiency. Household electricity usage can thus be categorized into two parts- (1) The minimum electricity a household uses as defined by the demand frontier, and (2) the excess part. The over-consumed or excess part indicates the potential savings that are possible within a household. Due to varying household characteristics, some households may consume more due to inefficiencies and over-use [25].

Eq. 2 assumes independently and identically distributed inefficiencies. It alone cannot find the potential factors that affect inefficiency. Different models were developed to identify the causal effects of inefficient energy use [71]. Battese and Coelli introduced the determinants of inefficiency in a vector $Zi$ to estimate the mean level of inefficiency [58]. The econometric formulation can now be expressed as Eq. 3. The inefficient part is assumed to have a normal distribution. $f(Z_i)$ contains variables of household electricity consumption that explain the inefficiency, $\delta$ is the vector of parameters to be estimated, and $\varepsilon_i$ represent the random errors which are assumed to be normally distributed $\varepsilon_i \sim \mathrm{N}^+ (0, \sigma^2)$. The variables should be chosen in such a way that they influence efficiency. It could include household socio-economic characteristics, equipment performance, occupants' behavior, etc.

$$f(Z_i) = \delta Z_i + \varepsilon_i$$
$$u_i \sim N^+(f(Z_i), \sigma_u^2) \quad (3)$$

Equations 2 and 3 can help evaluate how different household attributes affect inefficient electricity use and consumption frontiers. Further, the significant factors that impact inefficiencies in electricity consumption can also be identified. The model parameters are calculated using the maximum likelihood estimation (MLE) method.

For a family not operating on the frontier, the distance between the observed or actual consumption and the frontier represents inefficient use. A household's efficiency level can be expressed using Eq. 4, where Ei is the electricity consumed by a household and $E_i^F$ is the possible minimum consumption concerning the electricity demand frontier. The values of $TF_i$ range between 0 and 1. An efficiency value of 1 corresponds to a 100% efficient household, while 0 indicates a completely inefficient household. A score between 0 and 1 suggests some level of inefficiency. A low-efficiency score close to 0 thus implies high savings potential.

$$TF_i = {E_i^F}/{E_i} = e^{-u_i} \quad (4)$$

It is essential to highlight that before estimating the demand frontier, a diagnostic test, the likelihood ratio test (LR), is performed to confirm the presence of inefficiencies. The test compares the log-likelihood values of the Ordinary Least Squares (OLS) model with the SFA model, as shown in Eq 5. $L(H_0)$ represents the log-likelihood value of the OLS test and $L(H_1)$ denotes the log-likelihood values from the frontier model. The null hypothesis posits no substantial difference between the two models being compared, whereas the alternative hypothesis asserts that the compared model differs from the benchmark model. The presence or absence of inefficiency can also be determined by the value of lambda ($\lambda$), where $\lambda\ is\ {\sigma_u}/{\sigma_v}$. The model becomes a simple OLS if $\lambda$ is zero and vice versa.

$$LR = 2\,[L(H_o - L(H_1)] \quad (5)$$

STATA software was employed to run the SFA model. A total of 4 models, including OLS (model 1) for both slums and SRH, were run. Inefficiencies related to occupant behaviour were exempted from the first and second models. Model 2 is a half-normal (NHN) model without considering the effect of any determinants of inefficiency. It only separates noise and inefficiency terms. Log-likelihood values and λ were calculated for the OLS model and model 2 to confirm the presence of inefficiencies. The variables of inefficiency, like user behaviour in terms of the hours of using different appliances in addition to socio-economic characteristics, are introduced in the consequent models 3 and 4. Further, research evidence that households do not operate on the frontier due to differences in personal traits and the appliance stock [8]. Also, most households may not be clustered near the full efficiency and exhibit higher inefficiencies. Thus, model 4 is constructed to enable inefficiency determinants to impact the pre-truncation and allow the inefficiency distribution to have a nonzero mean. Hence, model 4 (TN) has $u_i \sim N^+(\mu_i, \sigma_u^2)$, where $\mu_i$ is the pre-truncation mean of the inefficiency term $u_i$. The model is the same as a half-normal (NHN) model for $\mu_i$= 0.

## 3.2 Case study area

Mumbai is an ideal location for examining electricity usage patterns in low-income housing, as 42% of its population resides in slum [72]. To provide access to basic amenities and improve the quality of life of the slum dwellers, the Government of Maharashtra launched the 'Slum Rehabilitation Housing' in 1995. The scheme entitles the slum dwellers to a tenured single-room apartment in high-rise buildings. The unit measures 250 sq. ft. and has an attached bathroom and kitchen. This shift, however, has been found to affect occupant behavior and, consequently, their uptake of appliances, as well as the demand for energy services. This study evaluates the efficiency of occupants' electricity usage behavior patterns for slums and Slum Rehabilitation Housing (SRH) in Mumbai. While slums are horizontally spread low-rise settlements (Figure 3), SRH refers to densely packed rehabilitated slum housing in a high-rise setup, as shown in Figure 4.

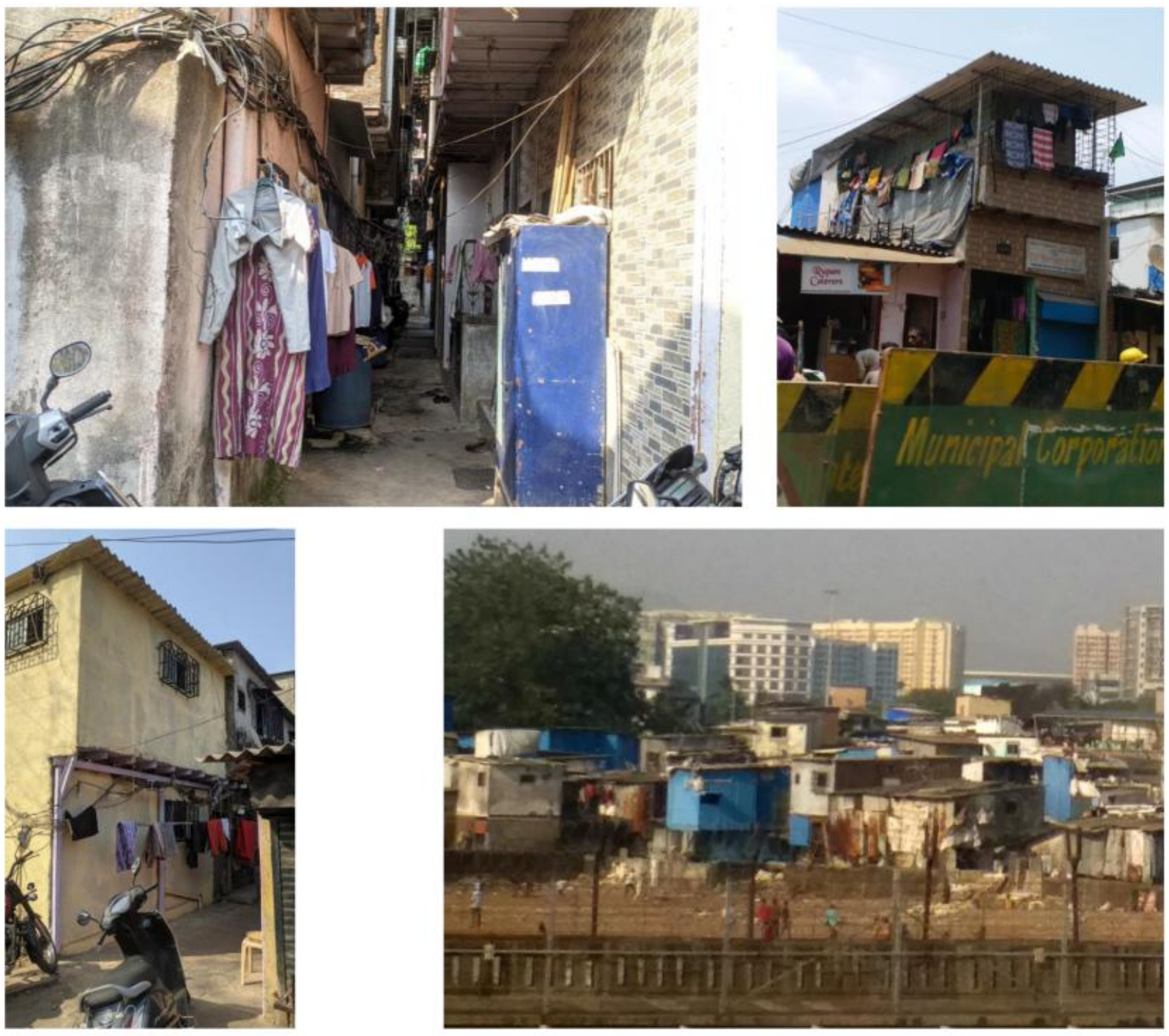

*Figure 3: Slum locations identified for household survey*

### 3.3 Unit of enquiry

The study focused on any adult household representatives, preferably females, present at the time of the survey. Since males spend a considerable time outdoors in the case of low-income housing, females who spend most of the time indoors could provide better understanding of appliance usage behavior.

### 3.4 Sample size determination

The study used data from 625 households across two slums and four SRHs in Mumbai, which were surveyed from March to April 2023. A minimum sample size of 385 was determined to represent the slum population, with a 95% confidence level and a 5% margin of error using Cochran's formula (1977). With an estimated 0.15 million slum dwellers being relocated to SRH, a minimal sample size of 384 was obtained for the reduced population sizes. Of the total 760 households surveyed, only 625 samples were found valid and selected for this study after the data cleaning (rejection of incomplete data) and sorting process.

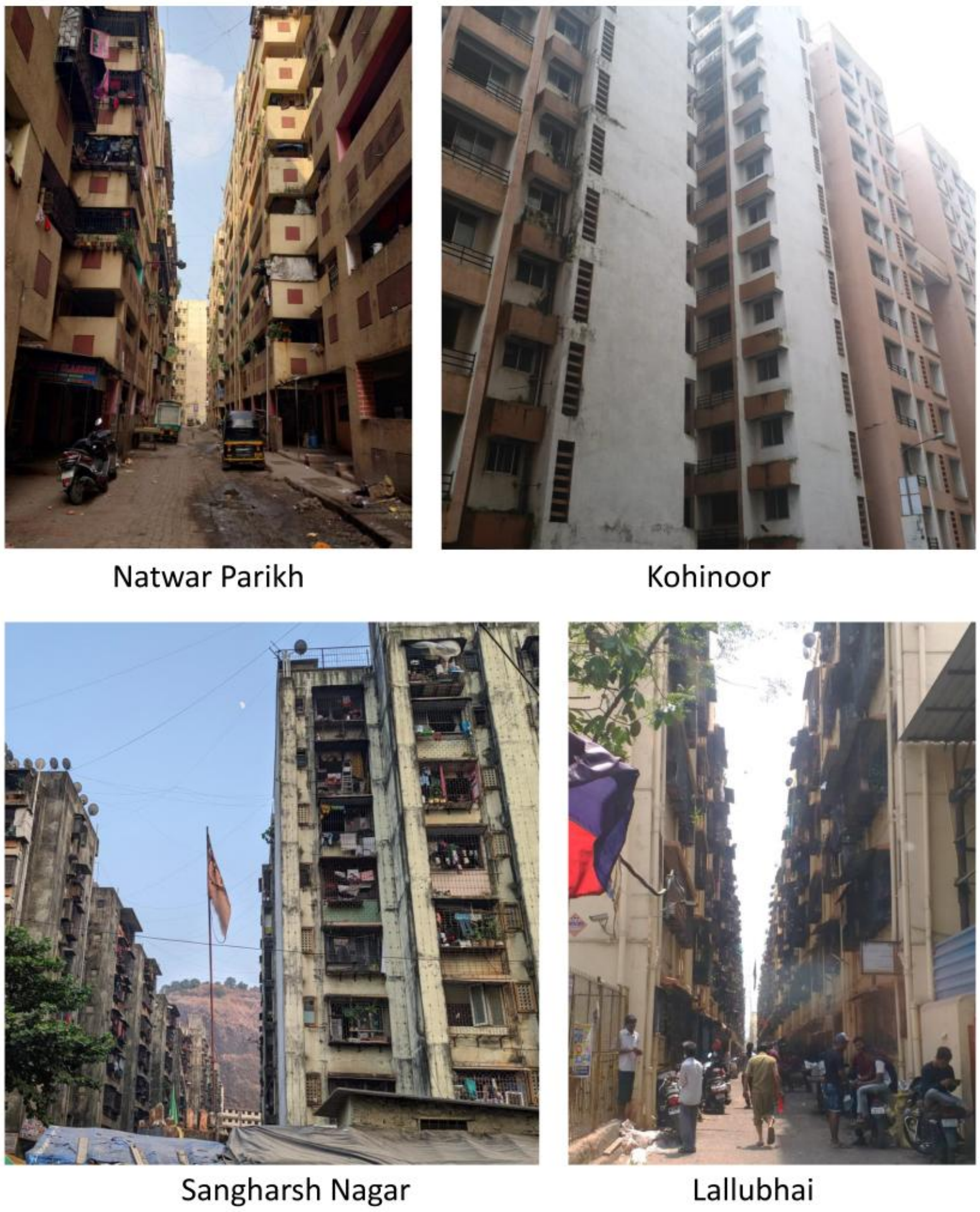

Figure 4: Slum rehabilitation housings selected for household survey

### 3.5 Sampling technique

The stratified random sampling approach was used to study the impact of household behaviour and technology on space electricity consumption in energy-poor households. As illustrated in Figure 3 and Figure 4, two slums and four typical resettlement colonies, respectively, were chosen for the survey. These included Ghatkopar and Govandi slums and Kohinoor SRH, Natwar Parikh, Sanghar Nagar SRH, and Lallubhai Compound. SRH fulfilling the following criteria were selected: i) the housing must be constructed between 2000 and 2010, and ii) it should be fully occupied for a minimum of 10-15 years.

### 3.6 Survey Design and data collection

Individual face-to-face discussions and interviews with household occupants were conducted at their respective slums and SRH to capture the household characteristics and applicable usage behavior. The questionnaire was organized in three languages- Hindi, English and Marathi. The participants were asked to complete the computer-aided personal interview (CAPI) based survey questionnaire. Prior to the actual data collection process, a pilot survey was held to test the structure and flow of questions, and the existence of any response biases. Additionally, the feedback collected on the pilot survey aided in strengthening the questionnaire by identifying the major appliances owned and used in such low-income housing. The questionnaire consisted of information pertaining to household characteristics household size, average household age, Work Force Participation Rate, income, etc.), appliances owned, occupant behavior (duration of performing different activities and hours of using different household appliances), electricity consumption, and any adaptive behavioral practices. Appliance ownership was coded as 0/1, where 1 indicates that a household owns the appliance and 0 if it does not. Table 2 provides a summary of the valid samples.

*Table 2: Statistical summary of the obtained household data (socio-economic characteristics, appliance ownership and usage) from slums and SRH in Mumbai*

***Housing type- SRH*** **(n= 412)**

| | Variable type | Mean | Std. Deviation | Min. | Max |
|---|---|---|---|---|---|
| **Dependent variable** | | | | | |
| Annual household electricity Consumption (kWh) | Continuous | 1583.800 | 556.881 | 351.00 | 3107.00 |
| **Household variables** | | | | | |
| WFPR | Continuous | 0.348 | 0.154 | 0 | 1 |
| HH Size | Continuous | 4.192 | 1.325 | 2 | 10 |
| Avg. HH age | Continuous | 31.574 | 8.544 | 13 | 65 |
| HH income | Ordinal (3,000 to 9,000 = quartile 1; 9,000 to 18,000 = quartile 2; 18,000 to 27,000 = quartile 3; 27,000 and above = quartile 4) | 2.939 | 0.737 | 1 | 4 |
| **Appliance ownership** | | | | | |
| Refrigerator | Dichotomous (No = 0; Yes = 1) | 0.9 | 0.3 | 0 | 1 |
| AC | Dichotomous (No = 0; Yes = 1) | 0.022 | 0.146 | 0 | 1 |
| Iron | Dichotomous (No = 0; Yes = 1) | 0.536 | 0.499 | 0 | 1 |
| Washing machine | Dichotomous (No = 0; Yes = 1) | 0.15 | 0.358 | 0 | 1 |
| Exhaust fan | Dichotomous (No = 0; Yes = 1) | 0.243 | 0.429 | 0 | 1 |
| TV | Dichotomous (No = 0; Yes = 1) | 0.939 | 0.239 | 0 | 1 |
| Laptop | Dichotomous (No = 0; Yes = 1) | 0.005 | 0.070 | 0 | 1 |
| Ceiling fan | Dichotomous (No = 0; Yes = 1) | 1 | 0.00 | 0 | 1 |
| Table fan | Dichotomous (No = 0; Yes = 1) | 0.036 | 0.188 | 0 | 1 |
| Mixer | Dichotomous (No = 0; Yes = 1) | 0.863 | 0.344 | 0 | 1 |
| CFL | Dichotomous (No = 0; Yes = 1) | 0.375 | 0.485 | 0 | 1 |
| LED | Dichotomous (No = 0; Yes = 1) | 0.957 | 0.203 | 0 | 1 |
| Bulb | Dichotomous (No = 0; Yes = 1) | 0.081 | 0.273 | 0 | 1 |
| **Hours of using appliances (daily in number of hours)** | | | | | |
| Refrigerator | Continuous | 21.612 | 7.193 | 0 | 24 |
| AC | Continuous | 0.056 | 0.4 | 0 | 4 |
| Iron | Continuous | 0.731 | 0.766 | 0 | 2 |
| Ceiling fan *(number of units is 2)* | Continuous | 26.454 | 10.637 | 7 | 46 |
| Table fan | Continuous | 0.357 | 1.935 | 0 | 15 |
| Washing machine | Continuous | 0.117 | 0.39 | 0 | 2 |
| Exhaust fan | Continuous | 0.34 | 0.67 | 0 | 4 |
| TV | Continuous | 4.964 | 2.334 | 0 | 10 |
| Laptop | Continuous | 0.01 | 0.139 | 0 | 2 |
| CFL | Continuous | 1.690 | 2.602 | 0 | 15 |
| LED *(number of units is more than 1)* | Continuous | 17.378 | 11.967 | 0 | 40 |
| Bulb | Continuous | 0.254 | 0.592 | 0 | 4 |

***Housing type- Slum*** **(n= 213)**

| | Variable type | Mean | Std. Deviation | Min. | Max. |
|---|---|---|---|---|---|
| **Dependent variable** | | | | | |
| Annual household electricity Consumption (kWh) | Continuous | 1612.908 | 593.892 | 351.00 | 3328.00 |
| **Household variables** | | | | | |
| WFPR | Continuous | 0.365 | 0.169 | 0 | 1 |
| HH Size | Continuous | 4.014 | 1.131 | 2 | 8 |
| Avg. HH age | Continuous | 33.627 | 8.886 | 14.25 | 62.5 |
| HH income | Ordinal (3,000 to 9,000 = quartile 1; 9,000 to 18,000 = quartile 2; 18,000 to 27,000 = quartile 3; 27,000 and above = quartile 4) | 2.878 | 0.815 | 1 | 4 |
| **Appliance ownership** | | | | | |
| Refrigerator | Dichotomous (No = 0; Yes = 1) | 0.817 | 0.388 | 0 | 1 |
| AC | Dichotomous (No = 0; Yes = 1) | 0.028 | 0.166 | 0 | 1 |
| Iron | Dichotomous (No = 0; Yes = 1) | 0.362 | 0.482 | 0 | 1 |
| Washing machine | Dichotomous (No = 0; Yes = 1) | 0.155 | 0.363 | 0 | 1 |
| Exhaust fan | Dichotomous (No = 0; Yes = 1) | 0.315 | 0.465 | 0 | 1 |
| TV | Dichotomous (No = 0; Yes = 1) | 0.962 | 0.191 | 0 | 1 |
| Laptop | Dichotomous (No = 0; Yes = 1) | 0.038 | 0.191 | 0 | 1 |
| Ceiling fan | Dichotomous (No = 0; Yes = 1) | 0.981 | 0.136 | 0 | 1 |
| Table fan | Dichotomous (No = 0; Yes = 1) | 0.056 | 0.231 | 0 | 1 |
| Mixer | Dichotomous (No = 0; Yes = 1) | 0.716 | 0.452 | 0 | 1 |
| CFL | Dichotomous (No = 0; Yes = 1) | 0.505 | 0.501 | 0 | 1 |
| LED | Dichotomous (No = 0; Yes = 1) | 0.971 | 0.169 | 0 | 1 |
| Bulb | Dichotomous (No = 0; Yes = 1) | 0.059 | 0.236 | 0 | 1 |
| **Hours of using appliances (daily in number of hours)** | | | | | |
| Refrigerator | Continuous | 19.606 | 9.304 | 0 | 24 |
| AC | Continuous | 0.08 | 0.539 | 0 | 5 |
| Iron | Continuous | 0.502 | 0.731 | 0 | 2 |
| Ceiling fan *(number of units is 2)* | Continuous | 22.033 | 11.609 | 0 | 48 |
| Table fan | Continuous | 0.512 | 2.523 | 0 | 18 |
| Washing machine | Continuous | 0.141 | 0.433 | 0 | 2 |
| Exhaust fan | Continuous | 0.446 | 0.754 | 0 | 4 |
| TV | Continuous | 4.211 | 1.673 | 0 | 10 |
| Laptop | Continuous | 0.042 | 0.263 | 0 | 2 |
| CFL | Continuous | 3.441 | 3.964 | 0 | 15 |
| LED *(number of units is more than 1)* | Continuous | 13.574 | 7.752 | 0 | 32 |
| Bulb | Continuous | 0.127 | 0.777 | 0 | 10 |

For this study, the SFA model used total annual electricity usage per household (kWh/household/yr) as the dependent variable to reflect the household's electricity consumption. The data pertaining to this output variable was calculated from the electricity bills obtained from the households during the field survey. Independent variables include socio-economic characteristics and occupant behaviour in terms of appliance ownership and hours of usage. Occupants' appliance usage behavior reflects whether or not the surveyed household runs an appliance.

As seen in Table 2, a high S.D. in the usage of AC and washing machines can be attributed to the heterogeneity in the households in terms of income, which restricts the consumption of energy-intensive appliances, as also observed by Ruijven et al. [73]. While some might run air conditioning frequently, others may limit usage to reduce high electricity costs. In the case of washing machines, a large disparity can also be attributed to the higher uptake among households with large WFPR owing to time constraints and enhanced convenience.

## 4. Results and Discussion

The value of the test statistic, LR, based on Eq 4 and calculated from OLS and Model 2 results for SRH and slum households, are 6.194 and 6.146. For a degree of freedom of 1 (as only $\sigma_u$ restricted), the critical value of the statistic at the 1% significance level is 5.412. Given that the obtained test statistic is 6.194, the null hypothesis of no technical inefficiency can be rejected. Further, the results from Model 2, as illustrated in Table 3 correspond to a non-zero value of lambda (λ) – 0.613 and 1.520, respectively, in SRH and slums. The standard deviations of the two error components are 0.347 and 0.213 in SRH and 0.291 and 0.443 in slums, respectively, for the idiosyncratic and inefficiency components, with a joint variance of 0.165 (in SRH) and 0.280 (in slums). These measures confirm the presence of inefficiency in SRH and slum households' electricity consumption.

The OLS model also provides a benchmark for the analysis, as it does not consider inefficiency, i.e., $u_i$=0. From the model outputs, it is found that owning an iron, mixer grinder, and AC significantly affects electricity consumption in SRH households. However, in slums, refrigerators, in addition to AC, are found to be the critical factors affecting electricity consumption. Intuitively, households with AC are likely to use more electricity, for instance, by 0.23 units in SRH (β=0.23, p-value=0.00) and 0.21 units in slums (β=0.211, p-value=0.009). The positive and significant coefficients for smaller appliances like iron (β=0.049, p-value=0.007) and mixer (β=0.048, p-value=0.053) in SRH can be ascribed to the greater time spent in homes, change in the employment patterns of women and consequently the increased electricity consumption [74]. No socioeconomic factors impacted electricity consumption, indicating significant variations in the SRH and slum households regarding consumption patterns and inefficiencies (Table 3 and Table 4). A positive and significant coefficient for refrigerators (β=0.114, p-value=0.002) in slums can be attributed to the prevalence of several informal small-scale home-based convenience stores, such as ice-cream shops or eateries dealing with perishable items, which often call for higher dependency on refrigeration.

*Table 3: Results of OLS and stochastic demand frontier function (NHN and TN models) for SRH*

| SRH | Model 1 (OLS) | | Model 2 (NHN without inefficiencies) | | Model 3 (NHN with inefficiencies) | | Model 4 (TN with inefficiencies) | |
|---|---|---|---|---|---|---|---|---|
| | Coefficient | p-value | Coefficient | p-value | Coefficient | p-value | Coefficient | p-value |
| ***Frontier*** | | | | | | | | |
| HH Income | 0.015 | 0.827 | 0.012 | 0.857 | 0.028 | 0.671 | 0.036 | 0.592 |
| HH size | 0.072 | 0.315 | 0.073 | 0.294 | 0.094 | 0.017 | -0.197 | 0.860 |
| Avg. HH age | -0.083 | 0.271 | -0.079 | 0.288 | -0.091 | 0.218 | -0.075 | 0.298 |
| WFPR | -0.060 | 0.287 | -0.061 | 0.265 | -0.153 | 0.037 | -0.070 | 0.190 |
| Exhaust | 0.006 | 0.741 | 0.006 | 0.759 | -0.004 | 0.090 | -0.001 | 0.947 |
| AC | 0.263 | 0.000 | 0.261 | 0.00 | 0.253 | 0.00 | 0.250 | 0.00 |
| Refrigerator | -0.019 | 0.531 | -0.020 | 0.503 | 0.131 | 0.002 | 0.000 | 0.986 |
| Washing machine | -0.009 | 0.677 | -0.008 | 0.702 | 0.030 | 0.252 | 0.037 | 0.214 |
| TV | 0.029 | 0.446 | 0.030 | 0.419 | 0.042 | 0.235 | 0.046 | 0.202 |
| Iron | 0.049 | 0.007 | 0.050 | 0.005 | 0.114 | 0.00 | 0.112 | 0.00 |
| Mixer | 0.048 | 0.053 | 0.049 | 0.044 | 0.043 | 0.067 | 0.034 | 0.144 |
| CFL | -0.009 | 0.634 | -0.010 | 0.610 | -0.074 | 0.004 | -0.149 | 0.002 |
| LED | 0.000 | 0.982 | 0.001 | 0.961 | 0.031 | 0.326 | 0.033 | 0.429 |
| Bulb | 0.012 | 0.716 | 0.010 | 0.739 | 0.031 | 0.00 | 0.026 | 0.506 |
| Constant | 8.187 | | 8.353 | | 8.576 | | 8.853 | |
| ***Inefficiencies*** | | | | | | | | |
| WFPR | | | | | -0.820 | 0.016 | | |
| Refrigerator _Hours of usage | | | | | 0.158 | 0.031 | 0.009 | 0.800 |
| Washing Machine_Hours of usage | | | | | 0.471 | 0.021 | 0.098 | 0.026 |
| Iron__Hours of usage | | | | | 0.809 | 0.021 | 0.136 | 0.014 |
| Bulb_Hours of usage | | | | | 0.743 | 0.015 | 2.08 | 0.880 |
| LED_Hours of usage | | | | | 0.282 | 0.121 | 0.055 | 0.179 |
| CFL_Hours of usage | | | | | -0.660 | 0.051 | -0.375 | 0.389 |
| **Log-likelihood value** | -178.853 | | -175.756 | | -182.778 | | -180.828 | |
| **Wald chi-square** | | | 42.65*** | | 84.13*** | | 63.82*** | |
| $\sigma_v$ | | | 0.347 | | 0.321 | | 0.319 | |
| $\sigma_u$ | | | 0.213 | | 1.427 | | 0.240 | |
| $\sigma^2 = \sigma_u^2 + \sigma_v^2$ | | | 0.165 | | 2.139 | | 0.566 | |
| $\lambda$ | | | 0.613 | | 4.445 | | 0.75 | |
| **Mean efficiency** | | | 0.85 | | 0.838 | | 0.77 | |

Note: Significant levels of 10%, 5%, and 1% are denoted by *, ** and *** respectively

The Wald chi-square statistic for all three models, as shown in Table 3 and Table 4, model 2 (42.65, p-value = 0.00; 60.72, p-value = 0.00), model 3 (84.13, p-value = 0.00; 77.29, p-value = 0.00), and model 4 (63.82, p-value = 0.00; 82.84, p-value = 0.00), in case of slums and SRH respectively, show that the chosen explanatory variables have a substantial importance in explaining the electricity usage in both slums and SRH. The log-likelihood values of different models can be used to select the model with a superior fit.

In the case of SRH, Model 3, or the normal-half normal model, depicts the highest log-likelihood values of -182.778. Thus, model 3 for SRH will be discussed in the consequent sections. For slums, however, both truncated normal and half-normal models show almost similar values for log-likelihood (Table 4); it was decided to move ahead with the half-normal model, considering the superior fit of the same model for SRH. The mean level of technical efficiency from Model 2, SRH, amounts to 0.85 and spans from 0.62 to 0.91 (Table 5). After controlling for the determinants of inefficiency, the score reduces to 0.83. The mean efficiency score in slums from the selected model 3 is 0.91, higher than SRH. A higher efficiency score in slums suggests that the factors considered in this study demonstrate substantial inefficiencies in SRH but in slums. A wide range in the estimated electricity efficiency in slums (S.D= 0. 149) compared to SRH (S.D= 0.126), as demonstrated in Table 5, indicates a large heterogeneity in slum households. This can be attributed to the variations in consumption patterns owing to economic practices like the presence of small-scale industries and appliances owned. Similar findings were reported in a recent study conducted by Dolšak et al. [42]. The authors highlight considerable heterogeneity amongst households regarding the stock of appliances used, the type of energy sources used, and demand patterns.

Figure 5 shows the estimated minimum electricity consumption for the surveyed households in slums and SRH. Here, the minimum consumption refers to the minimum (obtained from the frontier function) electricity that a household can consume to produce the same level of energy services, obtained at a higher (recorded) value. Of the total SRH households, 7% consume 20% or more electricity than is needed to produce the current energy services. In the case of slums, an estimated 15% of the total households consume 20% or more (extra) electricity than required to produce similar energy services. Around 4% of the total HHs waste at least 50% of the total consumed electricity, indicating immense potential for savings. A higher difference in slums is likely due to the small-scale household industries that could not be accounted for during the household surveys.

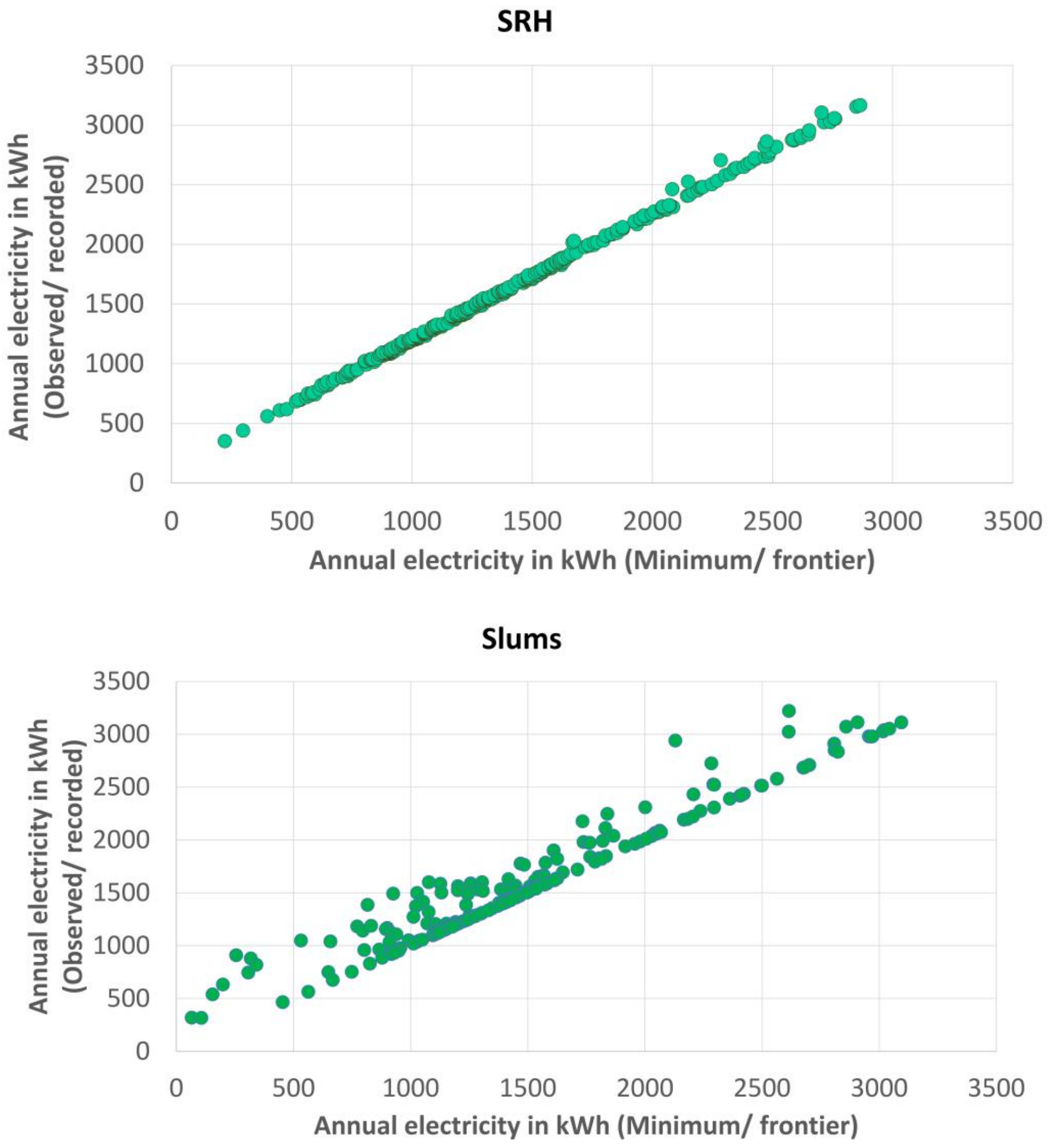

*Figure 5: Minimum vs observed electricity consumption (kWh) in SRH and slum households. Each point in the graph represents a household.*

The determinants of the minimum electricity consumption as defined by the frontier function are illustrated in Table 3 and Table 4. The output coefficients for ownership of appliances in the frontier function are positive, as the ownership would increase electricity use. Almost all appliances, except for LED, TV, and iron, as illustrated in Table 3, show a positive relationship with electricity consumption in SRH. A lack of association with TV indicates that the variable exhibits little or no variation. This could be likely due to the wide prevalence of the appliance in the study area. [23]. An insignificant value for iron ownership (p-value = 0.180) implies that even owning an iron does not guarantee its usage and thus does not affect electricity consumption or potential savings. As highlighted by other researchers, it is notable that owning an appliance does not indicate its usage, especially among low-income households [18]. An insignificant value for LED can be attributed to the equipment's low wattage.

For all other appliances, like refrigerator (β=0.131, p-value=0.002), exhaust fan (β=0.004, p-value=0.090), AC (β=0.253, p-value=0.00), lighting fixtures, and mixer (β=0.043, p-value=0.067), there exists a positive and significant relationship with electricity consumption. Other researchers also exhibit similar findings regarding a positive association with electricity demand [55]. Compared to other appliances, the ownership of refrigerators and ACs has the most significant coefficient, showing the highest positive effect on household electricity consumption. This implies that most households owning these appliances operate them frequently. Singh et al. point out that heavy home appliances such as refrigerators produce high electricity consumption in Indian urban households [75]. Refrigerators and AC have been reported to have a high statistically significant association with total residential electricity consumption [76]. A negative estimate for exhaust fans (β= -0.004, p-value=0.090) and CFL (β= -0.074, p-value=0.004) reveals an interesting observation. As these households often try to minimize their electricity usage owing to financial constraints, a reduction in electricity consumption due to exhaust fans and CFL is likely due to the tendency to not simultaneously operate other cooling and lighting devices like ceiling fans, table fans, AC, bulbs, etc., which otherwise consume more electricity owing to higher wattages [22]. An insignificant value for washing machines (p-value=0.252) highlights the residents' preference for manual laundering of clothes, as also observed in a recent study [23].

In the case of slums, appliances, in terms of ownership, that show a significant effect on electricity consumption include exhaust fans, ACs, refrigerators, irons, and mixers. A substantial and positive relationship for AC (β=0.365, p-value=0.000) can be attributed to the prevalence of several small-scale- industries, such as soap making, garments, etc., within slum houses that often require controlled indoor air conditions [18]. As discussed previously, a positive coefficient for refrigerators (β=0.115, p-value=0.000) can be attributed to shops dealing in confectionary items. Unlike SRH, slum households owning iron show a positive and statistically significant coefficient (β=0.107, p-value=0.020) with electricity consumption. This indicates that slum families owning iron often use the appliance. A different observation in the case of SRH can be attributed to the preference of higher income strata in SRH for commercial ironing services [23] (Table 3). A negative estimate for the exhaust fans (β= -0.058, p-value=0.049) suggests a similar user behavior as observed in SRH, i.e., the tendency to curb electricity usage. A non-significant estimate for washing machines is likely due to the practice of community living in slums, which often encourages everyday activities like washing and cleaning to be done together in outdoor spaces [18].

Out of the four socioeconomic variables, HH size and WFPR are statistically significant at a 1% level, indicating that the variables considerably impact household electricity consumption in SRH. The notable lack of association between electricity consumption and household income (p-value = 0.671) suggests that despite an increased HH income, families do not increase their electricity consumption or opt for more hours of appliance usage. This could be likely due to the large household sizes and, thus, the reduced per capita income. The findings are also supported by Zheng et al., who suggest that an increase in income in low-income households results in a rise in household survival needs like more energy-consuming foods [25]. These families shift to consumption upgrading, such as purchasing items after fulfilling their survival needs.

A statistically significant and positive relation with HH size (β= 0.094, p-value = 0.017) is in alignment with findings from other studies [48]. The consumption increases by 0.1 units for every additional member. The findings are also supported by existing literature that highlights that more occupants result in higher internal heat gain and, thus, more consumption [16]. However, some studies suggest that household size also has a scaling effect on electricity consumption [42]. More members could imply sharing resources such as lighting, cooling, etc., which could reduce total consumption. Further, when considering various types of household energy (for instance, fuel), due to the substitutional characteristics, the scale effect may surpass the marginal increasing effect caused by a rise in family size, thus leading to diminishing energy consumption [25].

WFPR exhibits a statistically significant and negative relationship (β= -0.153, p-value = 0.037) with electricity consumption. A negative relation indicates that families with a higher ratio of working members consume less electricity in SRH. While more working members are often associated with more significant household income, the current observations can be attributed to the lesser time spent inside homes amongst households with more working members [77].

Only WFPR is found to be statistically significant in slums. A positive estimate (β=0.113, p-value = 0.089) demonstrates that electricity consumption increases as the number of working members increases. Chen et al. made similar observations, suggesting that employment enables an increase in electricity consumption, particularly in low-income households [20]. The observation is likely due to the type of employment activity, majorly small-scale industries in which the residents are involved. In SRH, women, in particular, are engaged in work opportunities outside homes owing to the limited common spaces and their prohibited use for personal activities [18].

*Table 4: Results of OLS and stochastic demand frontier function (NHN and TN models) for slums*

| **Slum** | **Model 1 (OLS)** | | **Model 2** (NHN without inefficiencies) | | **Model 3** (NHN with inefficiencies) | | **Model 4** (TN with inefficiencies) | |
|---|---|---|---|---|---|---|---|---|
| | **Coefficient** | **p-value** | **Coefficient** | **p-value** | **Coefficient** | **p-value** | **Coefficient** | **p-value** |
| ***Frontier*** | | | | | | | | |
| HH Income | 0.082 | 0.432 | 0.077 | 0.423 | 0.043 | 0.623 | -0.031 | 0.710 |
| HH size | 0.104 | 0.30 | 0.089 | 0.393 | -0.003 | 0.975 | -0.392 | 0.041 |
| Avg. HH age | -0.046 | 0.605 | -0.077 | 0.509 | -0.080 | 0.447 | 0.508 | 0.033 |
| WFPR | 0.074 | 0.344 | 0.071 | 0.329 | 0.113 | 0.089 | 0.419 | 0.004 |
| Ceiling | 0.087 | 0.398 | 0.141 | 0.157 | 0.136 | 0.127 | 0.271 | 0.004 |
| Exhaust | -0.064 | 0.139 | -0.063 | 0.027 | -0.058 | 0.049 | -0.060 | 0.024 |
| AC | 0.211 | 0.009 | 0.225 | 0.002 | 0.365 | 0.000 | 0.211 | 0.001 |
| Refrigerator | 0.114 | 0.002 | 9.127 | 0.000 | 0.115 | 0.000 | 0.116 | 0.028 |
| Washing machine | -0.027 | 0.439 | -0.026 | 0.423 | 0.006 | 0.840 | -0.006 | 0.876 |
| TV | -0.014 | 0.846 | -0.019 | 0.780 | -0.008 | 0.901 | -0.028 | 0.633 |
| Iron | -0.000 | 0.979 | 0.010 | 0.686 | 0.107 | 0.020 | 0.426 | 0.00 |
| Mixer | 0.047 | 0.165 | 0.055 | 0.088 | 0.077 | 0.007 | 0.060 | 0.036 |
| CFL | 0.019 | 0.578 | 0.006 | 0.837 | 0.018 | 0.540 | 0.038 | 0.199 |
| LED | 0.109 | 0.190 | 0.127 | 0.107 | 0.097 | 0.202 | 0.126 | 0.055 |
| Bulb | 0.078 | 0.169 | 0.069 | 0.190 | 0.064 | 0.228 | 0.063 | 0.237 |
| Constant | 7.822 | | 8.393 | | 8.798 | | 8.653 | |
| ***Inefficiencies*** | | | | | | | | |
| HH size | | | | | -3.244 | 0.050 | -0.709 | 0.005 |
| Avg. HH age | | | | | - | - | 0.821 | 0.008 |
| WFPR | | | | | - | - | 0.476 | 0.024 |
| Exhaust_Hours of usage | | | | | -0.495 | 0.198 | - | - |
| Ceiling_Hours of usage | | | | | 0.513 | 0.367 | 0.090 | 0.131 |
| AC_Hours of usage | | | | | 2.677 | 0.043 | - | - |
| Refrigerator _Hours of usage | | | | | -0.947 | 0.304 | 0.011 | 0.751 |
| Washing Machine_Hours of usage | | | | | 0.825 | 0.096 | 0.031 | 0.646 |
| Iron__Hours of usage | | | | | 2.889 | 0.028 | 0.493 | 0.00 |
| Bulb_Hours of usage | | | | | -0.429 | 0.580 | - | - |
| LED_Hours of usage | | | | | -0.519 | 0.313 | - | - |
| CFL_Hours of usage | | | | | - | - | - | - |
| **Log-likelihood value** | -105.812 | | -102.739 | | -85.530 | | -86.193 | |
| **Wald chi-square** | | | 60.72*** | | 77.29*** | | 82.84*** | |
| $\sigma_v$ | | | 0.291 | | 0.333 | | 0.211 | |
| $\sigma_u$ | | | 0.443 | | 1.931 | | 0.377 | |
| $\sigma^2 = \sigma_u^2 + \sigma_v^2$ | | | 0.280 | | 3.839 | | 0.186 | |
| $\lambda$ | | | 1.520 | | 5.798 | | 1.786 | |
| **Mean efficiency** | | | 0.727 | | 0.911 | | 0.54 | |

Note: Significant levels of 10%, 5%, and 1% are denoted by *, ** and *** respectively

## 4.1 Determinants of electricity consumption efficiency

The latter part of Table 3 and Table 4 provide results on the determinants of inefficiency. It must be noted that the dependent variable is inefficiency; hence, a negative coefficient shows an improvement in efficiency, or a positive coefficient indicates enhanced inefficiency. Also, HH size, HH income, and average HH age were found to be insignificant in case of inefficiencies in SRH housing and were, therefore, eliminated from the final model.

The estimated efficiency values in SRH range from 0.188 to 0.987, with a mean value of 0.838 (Table 5 and normal curve in Figure 6a). It suggests that the inefficient use of household electricity consumption ranges from 1% to 81.2% of the total usage, with an average value of 16%. The blue bell-shaped curve, as demonstrated in Figure 6a, highlights that an estimated 40% of the households have inefficiencies higher than 16%. Thus, while some households are efficient in their usage, others towards the left have a high potential to decrease their consumption while maintaining the identical level of energy services currently derived. In slums, inefficient use ranges up to 79% of the total usage, with a mean value of around 9%. However, unlike SRH, only 30% of the total slum households report inefficiencies more significant than 9%, as visible in Figure 6b. The vast range of estimated electricity efficiency across SRH households, suggests a vast heterogeneity, which may be ascribed to diverse behavioural patterns and different appliance stock. As people shift from slums to tenured, permanent, and formal SRH, they experience a change in their social structure and mindset toward appliance purchases. The security of tenure gives them security and encourages the purchase of new appliances. Additionally, with changes in household practices like a shift of washing activity from community spaces, which were available in slums for performing activities like washing, to indoors, SRH HHs are more likely to buy convenience appliances like washing machines.

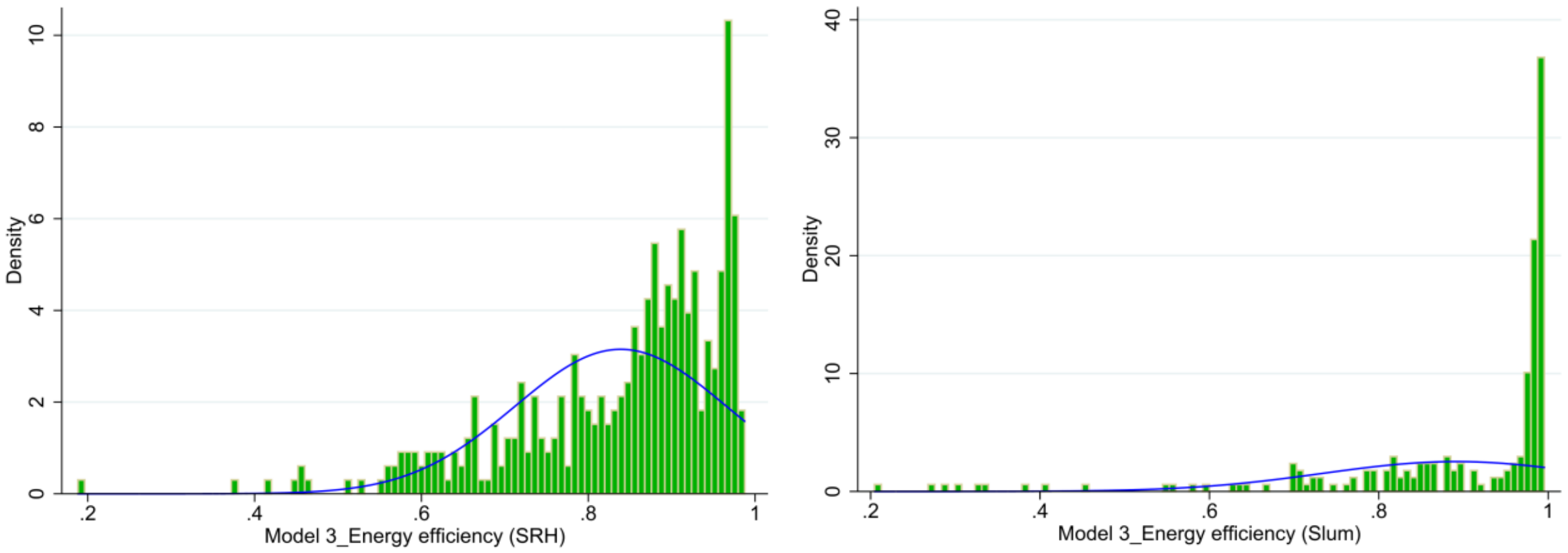

*Figure 6: Histogram of electricity efficiency index obtained from SFA model for (a) SRH (b) Slum*

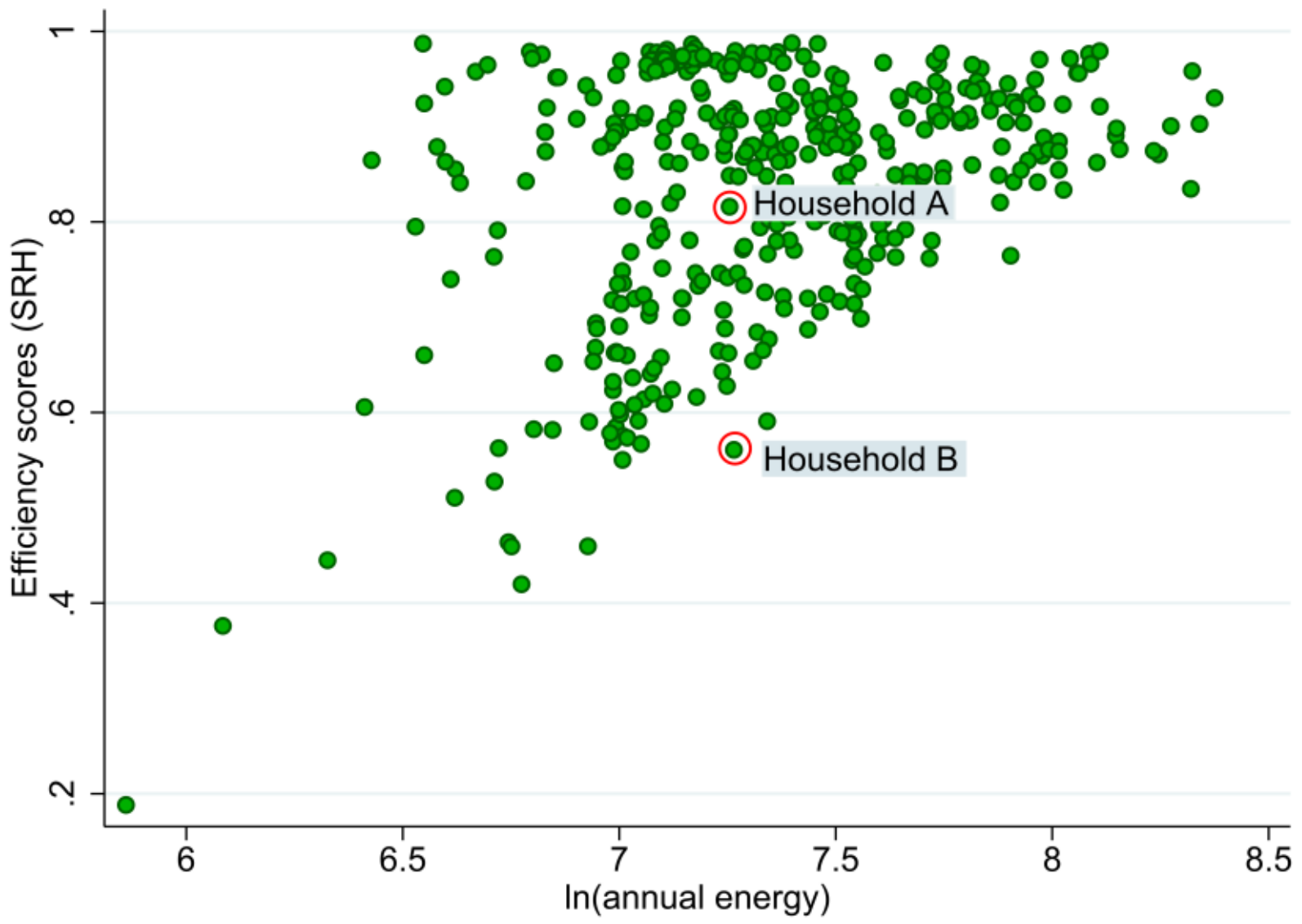

*Figure 7: Scatter plot of efficiency values and annual electricity consumption of the sampled SRH households.*

Figure 7 shows the scatter plot of efficiency scores obtained for SRH as a function of the annual electricity consumption of the sampled households. It is observed that several households with similar levels of consumption report different efficiency scores. For example, the consumption for households A and B are the same. However, the considerably different efficiency levels reflect the reliability of the results and the varying household characteristics.

The higher level of inefficiency in SRH (16.2%) than in slums (9%) as shown in Table 5, can be attributed to the higher appliance ownership amongst SRH residents [18]. Despite the similar socio-economic characteristics, families, after moving to SRH, purchase more appliances due to the permanency of tenure and aspiration to climb the social ladder. The change in daily practices, such as cooking meals and washing clothes, which are performed indoors in SRH, unlike slums, may also push families to purchase more appliances, such as washing machines. Due to financial constraints, these households also often buy older and less efficient appliances and equipment. Driven by social aspirations, they are more likely to obtain appliances at no cost from friends and family [21].

*Table 5: Distribution of efficiency scores in SRH and slums for the three analysed SFA models*

| | Mean | SD | Min. | Max. |
|---|---|---|---|---|
| **SRH** | | | | |
| **Model 2** | 0.85 | 0.037 | 0.621 | 0.918 |
| **Model 3** | 0.838 | 0.126 | 0.188 | 0.987 |
| **Model 4** | 0.773 | 0.171 | 0.310 | 0.993 |
| **Slum** | | | | |
| **Model 2** | 0.727 | 0.115 | 0.281 | 0.914 |
| **Model 3** | 0.911 | 0.149 | 0.205 | 0.998 |
| **Model 4** | 0.541 | 0.253 | 0.075 | 0.937 |

As shown in Table 3 and Table 4, a negative and statistically significant coefficient for WFPR (β= -0.820, p-value = 0.016) indicates that as the ratio of members working in a family in SRH increases, the inefficiencies in electricity consumption reduce. This could be likely due to the higher household income and, consequently, the presence of appliances with advanced electricity-efficient technology. Further, studies also establish a significant positive association between knowledge, awareness, and adoption of efficient technologies [78]. A higher household income is more likely to shift users into buying efficient appliances, owing to better education and knowledge [32]. Global studies also show a positive linkage between education level and intent to adopt efficiency measures [60].

The variable was, however, not found to yield any impact on inefficiencies in the case of slums. HH size considerably affects inefficiencies (β= -3.244, p-value = 0.50) in slums. A negative estimate highlights that as the HH size increases, inefficiency reduces. Zheng et al. observed a similar finding indicating that larger households reduce electricity-inefficient use [25]. The research findings can be ascribed to the reduced disposable household income with increased family size and, thus, fewer appliances owned. These households often prevent using appliances at all to mitigate electricity costs.

Hours of using different appliances, particularly refrigerator (β= 0.158, p-value = 0.031), washing machine (β= 0.471, p-value = 0.021), iron (β= 0.809, p-value = 0.021), CFL (β= -0.660, p-value = 0.051) and bulbs (β= 0.743, p-value = 0.015) are critical efficiency factors in SRH. In slums, appliances that show substantial effects on inefficiency include AC (β= 2.677, p-value = 0.043), washing machine (β= 0.825, p-value = 0.096), and iron (β= 2.889, p-value = 0.028). All these appliances (except for CFL) have a positive estimate indicating that their usage in slums and SRH households increases electricity inefficiency. This could be explained by the results observed in a recent study suggesting that low-income households own less electricity-efficient appliances, especially refrigerators and washing machines [20]. Further, since these families often own older appliances, the observed trends could be attributed to the age of the appliances. The results are in alignment with existing research that suggests that an increase in the average age of appliances lessens technical efficiency [68]. This is also supported by the empirical evidence that electricity efficiency labels for residential appliances are an essential energy conservation measure [79]. While existing research cites inefficiency in user behaviour, such as opening windows while using ACs or lack of electricity-saving consciousness, as a possible reason for inefficiency in electricity consumption [59], it is essential to highlight that these households often compromise their thermal comfort and convenience to reduce electricity costs [22].

Interestingly, unlike slums, hours of using refrigerators are a critical determinant of inefficiency in SRH. While households in SRH often switch off refrigerators, operate them for shorter durations [23] or may opt for older equipment to save on capital costs and overall power consumption; the option is not feasible in slums due to the employment activity type. Several slum households house small businesses or industries that require refrigeration.

The negative estimate for CFL reveals an interesting observation. Since it can be used instead of bulbs, which consume more electricity, increasing the hours of using LEDs and CFL lowers the total consumed electricity and brings down the inefficiency in consumption. Research also evidences that retrofitting with efficient lighting systems reports lowering residential electricity consumption and, consequently, cost savings [80]. However, an insignificant estimate for LED indicates its lower adoption among households. Thus, household appliances and specific household characteristics offer considerable opportunities to lessen electricity consumption in low-income families.

Twerefou et al. used a similar frontier model and found minimum efficiency scores of 66.3% for rural and urban households in Ghanaian households using similar input variables like household characteristics and type of appliances owned [55]. Weyman-Jones et al. found inefficiencies ranging between 4 and 43% using an input demand frontier function for Portuguese households with similar input variables [60]. Our results lie towards the lower end of the inefficiency levels obtained in these studies. The differences in estimated inefficiencies are likely to stem from differences in energy sources, variables employed, and methods used. For example, Weyman-Jones et al. estimate overall productive efficiency, composed of an allocative and technical efficiency, which departs from our research where only technical efficiency is calculated. Thus, direct comparisons of estimated efficiencies may not be possible. It is important to note that households are very diverse.

However, another recent research by Andor et al. used stochastic frontier analysis and disaggregated data from German households and found a mean inefficiency of around 20%, majorly due to HH size along with other variables [9]. For research-based in Slovenia, the authors estimated minimum inefficiency levels of around 23%, majorly due to the presence of inefficient capital stock and behavioral patterns [42]. The obtained determinants and inefficiency estimates reported by these authors in addition to Boogen align with the observed results of this study [8].

## 5. Conclusion

Studies on estimating efficiency in energy consumption have been carried out for countries like the US, Canada, China, and Japan, among others. However, little attention is paid to India's residential energy efficiency, in particular, the energy-poor families using household-level disaggregated data. Due to financial constraints, these households often compromise their thermal comfort, restrict their use of mechanical cooling, and use less efficient appliances. They are more likely to have older appliances that are less frequently retired from use, purchased from a second-hand market, or obtained at no cost from friends and family. Despite the long-term economic advantages, households often miss the potential for energy savings and witness adverse effects on health. Compounded by poor envelope design and indoor environmental conditions, the opportunity and the need to enhance efficiency is thus critical in energy-poor families.

Additionally, a lack of comprehensive data in India on energy consumption patterns often renders the strategies to enhance efficiency in social housing like slums and SRH uninformed. Gaining insight into their electricity consumption behaviour and subsequently estimating inefficiency is essential for improving their overall quality of life. The current study, therefore, examines the effect of household behaviour and technology on electricity consumption in energy-poor households for a selected case study area in Mumbai, India. Followed by an OLS model, it employs the stochastic frontier energy demand model to estimate the determinants of space electricity consumption and efficiency in slums and SRH based on data collected through household surveys. It adds to the current knowledge on household inefficiencies by quantitatively quantifying inefficiency for the first time in slums and SRH, predominantly in appliance consumption patterns.

The OLS and SFA model results reveal that the chosen explanatory (socio-economic and appliance updates) variables are crucial in explaining the electricity consumption and inefficiencies in the study area. The frontier functions exhibit a positive association between electricity consumption and most of the appliances owned in both housing types. Refrigerators and ACs have the most significant coefficient, indicative of their huge consumption values. This also suggests that most SRH households owning these appliances operate them frequently. In the case of slums, the obtained estimates can be attributed to the prevalence of several small-scale- industries that necessitate the need for controlled indoor air conditions and the preservation of perishable items. Interestingly, the results reveal that households often try to minimize their electricity usage by simultaneously not operating more than one cooling or lighting device. An insignificant relationship with washing machines highlights the residents' preference for manual laundering, a preferred practice in slums. Intuitively, HH size and WFPR are significant determinants of electricity consumption. The notable lack of association with income suggests that these energy-poor families curb their electricity consumption despite an increased HH income.

The SFA model further confirms the presence of inefficiency in SRH and slum households' electricity consumption. The estimated efficiency values range between 0.18 to 0.98 with a mean value of 0.83 in SRH based on the survey data. The scores vary between 0.20 to 0.99, with a mean value of 0.91 in slums. The results suggest the potential for reducing electricity consumption by at least 10% while maintaining the same level of energy services currently produced. SRH households are found to be more affected by inefficiencies than slums due to their higher appliance ownership. In the face of financial constraints driven by the security of tenure, increased time spent indoors, and social aspiration, purchasing older and less efficient appliances is prevalent in SRH. and contributes to higher inefficiencies. The presence of small-scale industries in slums results in heterogeneity in appliance usage behavior, and thus, a large SD in efficiency estimates is observed.

In terms of socio-economic variables, a negative association between inefficiency and WFPR highlights that with an increase in the ratio of working members, the inefficiencies in electricity consumption are reduced. A higher household income is more likely to shift users into buying efficient appliances, owing to better education and knowledge. A larger family size, on the contrary, is linked to lower inefficiencies likely due to diminished disposable household income and, thus, reduced usage of appliances to mitigate electricity costs. All the appliances (except for CFL) have been found to positively affect inefficiencies in electricity consumption. This strengthens the anecdotal evidence that low-income households own less electricity-efficient or older appliances and consequently compromise their thermal comfort and convenience to reduce electricity costs. Interestingly, unlike slums, refrigerator duration is a critical determinant of inefficiency in SRH. While SRH residents do not operate or operate them for shorter durations, the option is not feasible in slums due to small businesses requiring refrigeration. Overall, the obtained determinants and inefficiency estimates align with the results reported in the literature.

Notably, most of the existing studies have employed appliance ownership data along with household characteristics like HH size, income, age of the members, dwelling size, etc., to estimate inefficiencies in electricity consumption. This study, in addition to socio-economic variables, also takes into consideration the appliance ownership and usage data for the first time in social housing in India. While the obtained estimates are in line with the efficiency levels reported across the literature, it is important to note that any differences can be attributed to diversity in households, differences in residential energy sources, variables employed, and methodological frameworks used. Significant unobserved heterogeneity may exist that this paper could not account for. However, this can be solved by using panel data in future research work.

The study helps quantify the potential for electricity savings in energy-poor households by maintaining the same level of services currently produced. The results provide the information about households' lifestyles needed for informing effective demand side management programs. The findings quantify the existing inefficiencies to guide energy policies. These policies should be taken up in unison with the design of social housing like slums and SRH and appliance purchase behaviour. This study recommends a set of policy measures for improving the overall quality of life of energy-poor households through improved thermal comfort and access to energy-efficient appliances-

- The findings contribute to identifying drivers of inefficiencies in electricity consumption in developing countries. This difference between the observed mean efficiency levels between slums and SRH indicates a fundamental difference in appliance ownership patterns within the two typologies. A high value of electricity inefficiency highlights an immense opportunity to implement electricity efficiency measures while providing the same energy services, especially in SRH.

- The empirical evidence of the existence of inefficiencies highlights the need for investing in and supporting household energy management and promoting access to vital and efficient energy services. The government can create ration shops to provide subsidies for households to purchase essential (appliances with higher utility) energy-efficient electric appliances. Price subsidies can encourage households to upgrade to modern energy-efficient appliances and reduce their wasted electricity while improving their living conditions. The use of intelligent meters can be promoted to help households perceive changes in electricity consumed and thus operate essential appliances for fulfilling basic daily needs like refrigeration and cooling.
- Results show that electricity-saving opportunities are associated with characteristics such as awareness of energy-saving practices, preferences in choosing efficient equipment, the age of installed appliances, and the quality of dwelling. Thus, suitable measures like awareness campaigns can be organized to spread awareness and educate households on technological advancement and the need to purchase star-rated appliances, as the lifetime cost of operating an inefficient appliance is higher than an efficient one. Additionally, the effects of inefficient appliances on inhabitants' long-term health and well-being should be discussed as these households often try to curb their hours of operating certain appliances.
- Improving the energy performance of new buildings and accelerating the energy efficiency retrofit of existing residential buildings can reduce cooling needs. Reflective coatings can be applied to envelopes to reduce heat gains and maintain a comfortable indoor environment.
- As these families refrain from using mechanical cooling devices to reduce electricity bills, the housing units could be designed such that there is appropriate air flow within. Firstly, a careful consideration of the positioning and design of fenestration, in addition to adopting cross-style layouts, can further enhance ventilation and improve indoor thermal comfort. Second, as large household sizes result in higher internal heat gains, the government can allocate units based on a standard HH size range and allow occupancy-based changes in design to improve indoor environment quality. Third, the first few electricity units can be provided free of cost to these households to enhance their thermal comfort and reduce financial stress.

Overall, this study offers a comprehensive analysis of household electricity usage along with its income and non-income determinants. These results hold implications for energy policymakers, aiding in formulating incentives and strategies, particularly targeting low-income urban housing. However, there are a few limitations of this study. The current study focuses explicitly on electricity efficiency in terms of appliance usage behaviour and household characteristics. It did not examine inefficiencies from other energy-consuming activities like cooking and heating. Future research can widen the scope to include other sources of energy in addition to expanding the focus to integrate other climatic zones where heating is prevalent. It can also include other regions with typical socio-economic and cultural contexts like clothing and food habits.

Since there exists considerable heterogeneity in slums and SRH households, future investigation can find the specific electricity-saving potentials for different income-level households within such social housing. The existing survey-based data can be validated using measured data collected via smart energy meters installed in typical housing units. This research may have certain unobserved variables that were not accounted for. Future studies can overcome this issue by utilizing the SFA model based on panel data. Collecting time-series data on electricity consumption patterns and corresponding behavioural can provide interesting observations in terms of variations in consumption and efficiency patterns, if any. In the future, it would be interesting to integrate panel data, particularly relating to seasonal variations, and opening schedules of windows and doors to improve energy savings. The study can also be extended to the national level by utilizing data on appliance ownership from government sources such as NSSO (National Sample Survey Office).

## Acknowledgments

The authors would like to thank the journals that permitted the use of Figure 2.